


%
%


\def\famname{
 \textfont0=\textrm \scriptfont0=\scriptrm
 \scriptscriptfont0=\sscriptrm
 \textfont1=\textmi \scriptfont1=\scriptmi
 \scriptscriptfont1=\sscriptmi
 \textfont2=\textsy \scriptfont2=\scriptsy \scriptscriptfont2=\sscriptsy
 \textfont3=\textex \scriptfont3=\textex \scriptscriptfont3=\textex
 \textfont4=\textbf \scriptfont4=\scriptbf \scriptscriptfont4=\sscriptbf
 \skewchar\textmi='177 \skewchar\scriptmi='177
 \skewchar\sscriptmi='177
 \skewchar\textsy='60 \skewchar\scriptsy='60
 \skewchar\sscriptsy='60
 \def\rm{\fam0 \textrm} \def\bf{\fam4 \textbf}}
\def\sca#1{scaled\magstep#1} \def\scah{scaled\magstephalf} 
\def\twelvepoint{
 \font\textrm=cmr12 \font\scriptrm=cmr8 \font\sscriptrm=cmr6
 \font\textmi=cmmi12 \font\scriptmi=cmmi8 \font\sscriptmi=cmmi6 
 \font\textsy=cmsy10 \sca1 \font\scriptsy=cmsy8
 \font\sscriptsy=cmsy6
 \font\textex=cmex10 \sca1
 \font\textbf=cmbx12 \font\scriptbf=cmbx8 \font\sscriptbf=cmbx6
 \font\it=cmti12
 \font\sectfont=cmbx12 \sca1
 \font\refrm=cmr10 \scah \font\refit=cmti10 \scah
 \font\refbf=cmbx10 \scah
 \def\twelverm{\textrm} \def\twelveit{\it} \def\twelvebf{\textbf}
 \famname \textrm 
 \voffset=.04in \hoffset=.21in
 \normalbaselineskip=18pt plus 1pt \baselineskip=\normalbaselineskip
 \parindent=21pt
 \setbox\strutbox=\hbox{\vrule height10.5pt depth4pt width0pt}}


\catcode`@=11

{\catcode`\'=\active \def'{{}^\bgroup\prim@s}}

\def\screwcount{\alloc@0\count\countdef\insc@unt}   
\def\screwdimen{\alloc@1\dimen\dimendef\insc@unt} 
\def\screwbox{\alloc@4\box\chardef\insc@unt}

\catcode`@=12


\overfullrule=0pt			
\voffset=.04in \hoffset=.21in
\vsize=9in \hsize=6in
\parskip=\medskipamount	
\lineskip=0pt				
\normalbaselineskip=18pt plus 1pt \baselineskip=\normalbaselineskip
\abovedisplayskip=1.2em plus.3em minus.9em 
\belowdisplayskip=1.2em plus.3em minus.9em	
\abovedisplayshortskip=0em plus.3em	
\belowdisplayshortskip=.7em plus.3em minus.4em	
\parindent=21pt
\setbox\strutbox=\hbox{\vrule height10.5pt depth4pt width0pt}
\def\makefootline{\baselineskip=30pt \line{\the\footline}}
\footline={\ifnum\count0=1 \hfil \else\hss\twelverm\folio\hss \fi}
\pageno=1


\def\boxit#1{\leavevmode\thinspace\hbox{\vrule\vtop{\vbox{
	\hrule\kern1pt\hbox{\vphantom{\bf/}\thinspace{\bf#1}\thinspace}}
	\kern1pt\hrule}\vrule}\thinspace}
\def\Boxit#1{\noindent\vbox{\hrule\hbox{\vrule\kern3pt\vbox{
	\advance\hsize-7pt\vskip-\parskip\kern3pt\bf#1
	\hbox{\vrule height0pt depth\dp\strutbox width0pt}
	\kern3pt}\kern3pt\vrule}\hrule}}


\def\put(#1,#2)#3{\screwdimen\unit  \unit=1in
	\vbox to0pt{\kern-#2\unit\hbox{\kern#1\unit
	\vbox{#3}}\vss}\nointerlineskip}

%
%
%
%
%
%
%

\def\\{\hfil\break}

\def\center{\leftskip=0pt plus 1fill \rightskip=\leftskip \parindent=0pt
 \def\textindent##1{\par\hangindent21pt\footrm\noindent\hskip21pt
 \llap{##1\enspace}\ignorespaces}\par}
\def\unnarrower{\leftskip=0pt \rightskip=\leftskip}
\def\thetitle#1#2#3#4#5{
 \font\titlefont=cmbx12 \sca2 \font\footrm=cmr10 \font\footit=cmti10
  \twelverm
	{\hbox to\hsize{#4 \hfill ITP-SB-#3}}\par
	\vskip.8in minus.1in {\center\baselineskip=1.44\normalbaselineskip
 {\titlefont #1}\par}{\center\baselineskip=\normalbaselineskip
 \vskip.5in minus.2in #2
	\vskip1.4in minus1.2in {\twelvebf ABSTRACT}\par}
 \vskip.1in\par
 \narrower\par#5\par\unnarrower\vskip3.5in minus2.3in\eject}
\def\paper\par#1\par#2\par#3\par#4\par#5\par{\twelvepoint
	\thetitle{#1}{#2}{#3}{#4}{#5}} 
\def\author#1#2{#1 \vskip.1in {\twelveit #2}\vskip.1in}
\def\ITP{Institute for Theoretical Physics\\
	State University of New York, Stony Brook, NY 11794-3840}
\def\WS{W. Siegel\footnote{${}^1$}{       
 Internet address: siegel@insti.physics.sunysb.edu.}}


\def\sect#1\par{\par\ifdim\lastskip<\medskipamount
	\bigskip\medskip\goodbreak\else\nobreak\fi
	\noindent{\sectfont{#1}}\par\nobreak\medskip} 
\def\itemize#1 {\item{[#1]}}	
\def\vol#1 {{\refbf#1} }		 

\def\ref#1{\setbox0=\hbox{M}$\vbox to\ht0{}^{#1}$}


\def\NP #1 {{\refit Nucl. Phys.} {\refbf B{#1}} }
\def\PL #1 {{\refit Phys. Lett.} {\refbf{#1}} }
\def\PR #1 {{\refit Phys. Rev. Lett.} {\refbf{#1}} }
\def\PRD #1 {{\refit Phys. Rev.} {\refbf D{#1}} }


\hyphenation{pre-print}
\hyphenation{quan-ti-za-tion}

%
%

\def\on#1#2{{\buildrel{\mkern2.5mu#1\mkern-2.5mu}\over{#2}}}
\def\dt#1{\on{\hbox{\bf .}}{#1}}                
\def\ddt#1{\on{\hbox{\bf .\kern-1pt.}}#1}    
\def\slap#1#2{\setbox0=\hbox{$#1{#2}$}
	#2\kern-\wd0{\hbox to\wd0{\hfil$#1{/}$\hfil}}}
\def\sla#1{\mathpalette\slap{#1}}                
\def\bop#1{\setbox0=\hbox{$#1M$}\mkern1.5mu
	\vbox{\hrule height0pt depth.04\ht0
	\hbox{\vrule width.04\ht0 height.9\ht0 \kern.9\ht0
	\vrule width.04\ht0}\hrule height.04\ht0}\mkern1.5mu}
\def\bo{{\mathpalette\bop{}}}                        
\def~{\widetilde} 
\mathcode`\*="702A                  
\def\in{\relax\ifmmode\mathchar"3232\else{\refit in\/}\fi} 
\def\f#1#2{{\textstyle{#1\over#2}}}	   
\def\half{{\textstyle{1\over{\raise.1ex\hbox{$\scriptstyle{2}$}}}}}

\catcode`\^^?=13				    
\catcode128=13 \def €{\"A}                 
\catcode129=13 \def {\AA}                 
\catcode130=13 \def '{\c}           	   
\catcode131=13 \def ƒ{\'E}                   
\catcode132=13 \def "{\~N}                   
\catcode133=13 \def …{\"O}                 
\catcode134=13 \def †{\"U}                  
\catcode135=13 \def ‡{\'a}                  
\catcode136=13 \def ˆ{\`a}                   
\catcode137=13 \def ‰{\^a}                 
\catcode138=13 \def Š{\"a}                 
\catcode139=13 \def ‹{\~a}                   
\catcode140=13 \def Œ{\alpha}            
\catcode141=13 \def {\chi}                
\catcode142=13 \def Ž{\'e}                   
\catcode143=13 \def {\`e}                    
\catcode144=13 \def {\^e}                  
\catcode145=13 \def '{\"e}                
\catcode146=13 \def '{\'\i}                 
\catcode147=13 \def "{\`\i}                  
\catcode148=13 \def "{\^\i}                
\catcode149=13 \def •{\"\i}                
\catcode150=13 \def –{\~n}                  
\catcode151=13 \def —{\'o}                 
\catcode152=13 \def ˜{\`o}                  
\catcode153=13 \def ™{\^o}                
\catcode154=13 \def š{\"o}                 
\catcode155=13 \def ›{\~o}                  
\catcode156=13 \def œ{\'u}                  
\catcode157=13 \def {\`u}                  
\catcode158=13 \def ž{\^u}                
\catcode159=13 \def Ÿ{\"u}                
\catcode160=13 \def  {\tau}               
\catcode161=13 \mathchardef ¡="2203     
\catcode162=13 \def ¢{\oplus}           
\catcode163=13 \def £{\relax\ifmmode\to\else\itemize\fi} 
\catcode164=13 \def ¤{\subset}	  
\catcode165=13 \def ¥{\infty}           
\catcode166=13 \def ¦{\mp}                
\catcode167=13 \def §{\sigma}           
\catcode168=13 \def ¨{\rho}               
\catcode169=13 \def ©{\gamma}         
\catcode170=13 \def ª{\leftrightarrow} 
\catcode171=13 \def «{\relax\ifmmode\acute\else\expandafter\'\fi}
\catcode172=13 \def ¬{\relax\ifmmode\expandafter\ddt\else\expandafter\"\fi}
\catcode173=13 \def ­{\equiv}            
\catcode174=13 \def ®{\approx}          
\catcode175=13 \def ¯{\Omega}          
\catcode176=13 \def °{\otimes}          
\catcode177=13 \def ±{\ne}                 
\catcode178=13 \def ²{\le}                   
\catcode179=13 \def ³{\ge}                  
\catcode180=13 \def ´{\upsilon}          
\catcode181=13 \def µ{\mu}                
\catcode182=13 \def ¶{\delta}             
\catcode183=13 \def ·{\epsilon}          
\catcode184=13 \def ¸{\Pi}                  
\catcode185=13 \def ¹{\pi}                  
\catcode186=13 \def º{\beta}               
\catcode187=13 \def »{\partial}           
\catcode188=13 \def ¼{\nobreak\ }       
\catcode189=13 \def ½{\zeta}               
\catcode190=13 \def ¾{\sim}                 
\catcode191=13 \def ¿{\omega}           
\catcode192=13 \def À{\dt}                     
\catcode193=13 \def Á{\gets}                
\catcode194=13 \def Â{\lambda}           
\catcode195=13 \def Ã{\nu}                   
\catcode196=13 \def Ä{\phi}                  
\catcode197=13 \def Å{\xi}                     
\catcode198=13 \def Æ{\psi}                  
\catcode199=13 \def Ç{\int}                    
\catcode200=13 \def È{\oint}                 
\catcode201=13 \def É{\relax\ifmmode\cdot\else\vol\fi}    
\catcode202=13 \def Ê{\relax\ifmmode\,\else\thinspace\fi}
\catcode203=13 \def Ë{\`A}                      
\catcode204=13 \def Ì{\~A}                      
\catcode205=13 \def Í{\~O}                      
\catcode206=13 \def Î{\Theta}              
\catcode207=13 \def Ï{\theta}               
\catcode208=13 \def Ð{\relax\ifmmode\bar\else\expandafter\=\fi}
\catcode209=13 \def Ñ{\overline}             
\catcode210=13 \def Ò{\langle}               
\catcode211=13 \def Ó{\relax\ifmmode\{\else\ital\fi}      
\catcode212=13 \def Ô{\rangle}               
\catcode213=13 \def Õ{\}}                        
\catcode214=13 \def Ö{\sla}                      
\catcode215=13 \def ×{\relax\ifmmode\check\else\expandafter\v\fi}
\catcode216=13 \def Ø{\"y}                     
\catcode217=13 \def Ù{\"Y}  		    
\catcode218=13 \def Ú{\Leftarrow}       
\catcode219=13 \def Û{\Leftrightarrow}       
\catcode220=13 \def Ü{\relax\ifmmode\Rightarrow\else\sect\fi}
\catcode221=13 \def Ý{\sum}                  
\catcode222=13 \def Þ{\prod}                 
\catcode223=13 \def ß{\widehat}              
\catcode224=13 \def à{\pm}                     
\catcode225=13 \def á{\nabla}                
\catcode226=13 \def â{\quad}                 
\catcode227=13 \def ã{\in}               	
\catcode228=13 \def ä{\star}      	      
\catcode229=13 \def å{\sqrt}                   
\catcode230=13 \def æ{\^E}			
\catcode231=13 \def ç{\Upsilon}              
\catcode232=13 \def è{\"E}    	   	 
\catcode233=13 \def é{\`E}               	  
\catcode234=13 \def ê{\Sigma}                
\catcode235=13 \def ë{\Delta}                 
\catcode236=13 \def ì{\Phi}                     
\catcode237=13 \def í{\`I}        		   
\catcode238=13 \def î{\iota}        	     
\catcode239=13 \def ï{\Psi}                     
\catcode240=13 \def ð{\times}                  
\catcode241=13 \def ñ{\Lambda}             
\catcode242=13 \def ò{\cdots}                
\catcode243=13 \def ó{\^U}			
\catcode244=13 \def ô{\`U}    	              
\catcode245=13 \def õ{\bo}                       
\catcode246=13 \def ö{\relax\ifmmode\hat\else\expandafter\^\fi}
\catcode247=13 \def÷{\relax\ifmmode\tilde\else\expandafter\~\fi}
\catcode248=13 \def ø{\ll}                         
\catcode249=13 \def ù{\gg}                       
\catcode250=13 \def ú{\eta}                      
\catcode251=13 \def û{\kappa}                  
\catcode252=13 \def ü{\half}     		 
\catcode253=13 \def ý{\Gamma} 		
\catcode254=13 \def þ{\Xi}   			
\catcode255=13 \def ÿ{\relax\ifmmode{}^{\dagger}{}\else\dag\fi}


\def\ital#1Õ{{\it#1\/}}	     
\def\un#1{\relax\ifmmode\underline#1\else $\underline{\hbox{#1}}$
	\relax\fi}

\def\tdt#1{\on{\hbox{\bf .\kern-1pt.\kern-1pt.}}#1}   
\def\({\eqno(}

\def\refs{\sect{REFERENCES}\par\medskip \frenchspacing 
	\parskip=0pt \refrm \baselineskip=1.23em plus 1pt
	\def\ital##1Õ{{\refit##1\/}}}


\def\õ#1{
	\screwcount\num
	\num=1
	\screwdimen\downsy
	\downsy=-1.5ex
	\mkern-3.5mu
	õ
	\loop
	\ifnum\num<#1
	\llap{\raise\num\downsy\hbox{$õ$}}
	\advance\num by1
	\repeat}
\def\upõ#1#2{\screwcount\numup
	\numup=#1
	\advance\numup by-1
	\screwdimen\upsy
	\upsy=.75ex
	\mkern3.5mu
	\raise\numup\upsy\hbox{$#2$}}


\catcode`\|=\active \catcode`\<=\active \catcode`\>=\active 
\def|{\relax\ifmmode\delimiter"026A30C \else$\mathchar"026A$\fi}
\def<{\relax\ifmmode\mathchar"313C \else$\mathchar"313C$\fi}
\def>{\relax\ifmmode\mathchar"313E \else$\mathchar"313E$\fi}


\paper

T-DUALITY INVARIANCE IN\\ RANDOM LATTICE STRINGS

\author\WS\ITP

96-6

March 6, 1996

Preserving the T-duality invariance of the continuum string in its random
lattice regularization uniquely determines the random matrix model
potential.  For D=0 the duality transformation can be performed explicitly
on the matrix action, and replaces color with flavor; invariance thus
requires that the color and flavor groups be the same.

Ü1.  INTRODUCTION

By definition renormalization must preserve as many properties as
possible of a classical field theory.  In particular this includes a maximal
set of nonanomalous global or local symmetries.  The simplest way to
renormalize consistently with this requirement is to use a regularization
scheme that automatically preserves as many as possible of these
symmetries, so as to avoid the inconvenience of finite renormalizations
and Ward-Takahashi-Slavnov-Taylor identities.  Regularization schemes
that preserve symmetries (such as dimensional regularization) also tend
to make calculations simpler.

One regularization scheme that has the added benefit of providing a
calculational scheme for nonperturbative quantum field theory is the use
of a lattice.  In the case of strings, the worldsheet is replaced with a
lattice, whose two-dimensional nature is preserved by the topological
properties of the $1/N$ expansion, and whose arbitrary worldsheet metric
is manifested through the randomness of the Feynman diagrams that
represent the lattices [1].  These Feynman diagrams belong to an
underlying field theory whose bound states are strings.  Generally the
continuum limit, which is the limit of vanishing worldsheet cosmological
constant, also requires the limit of infinite colors ($N$), as a result of
dimensional transmutation.  However, the ``lumpy" string that results
from fixed $N$ may be of more physical relevance, and can be treated by
expansion about the continuum string.

The fact that this lattice replaces the worldsheet, and not physical
spacetime (as in lattice QCD), means that spacetime symmetries, such as
Poincar«e invariance, can be manifestly preserved by this regularization. 
However, one spacetime symmetry of (closed) string theory that has been
completely ignored on the worldsheet lattice is T-duality invariance [2]: 
Although T-duality transformations on random lattices have been
considered [3], the fact that the continuum string theory is ÓinvariantÕ
under this transformation has never been used to define the lattice
regularization of the worldsheet.  By T-duality we refer here to the
transformation $»_m x = ·_{mn}»^n x'$; i.e., Hodge duality on $»x$.  In this
paper T-duality will be applied to the underlying fundamental fields of the
random matrix model, and not just the composite fields of the string
theory, whose T-duality transformations are commonly studied through
the introduction of background fields.

Since neither the regularization nor the renormalization of these matrix
models has been shown to preserve T-duality invariance, it is not clear
whether spurious terms might have been introduced into the path
integral, in the same way that using a momentum-space cutoff for QCD
produces terms that aren't gauge invariant.  Although such terms might be
avoided in lower dimensions, universality is not expected in
more-physical cases.  Also, we may want to preserve symmetries for
finite $N$.  In this paper we show that duality invariance uniquely
determines the lattice regularization by fixing the matrix model potential,
and find the resulting model.  At least in D=0, this model of ``gluons"
(hermitian matrices) has a natural interpretation in terms of a theory of
self-interacting bosonic ``quarks" (complex matrices), whose action has
just quadratic and quartic terms.

In D=0 the only effect of the duality transformation is to replace a
Feynman graph with the dual graph.  We show how to explictly perform
this transformation on the matrix model action.  Duality switches the
color and flavor of the quarks, so duality invariance requires the equality
of the color and flavor groups.

Ü2.  THE MODEL

A T-duality transformation on a quantum field theory, as defined by
treating that theory as the underlying theory of a string theory, has two
main effects on the Feynman diagrams of that field theory [4]:  (1) It
replaces the propagator with its Fourier transform.  This is a trivial
invariance for the usual strings, where these propagators are Gaussians. 
(However, in more-realisitic field theories, with $1/p^2$ propagators, this
is an invariance only in D=4 [5].)  (2) It replaces the Feynman diagram
(lattice) with the dual lattice --- the graph resulting from replacing
vertices with loops (faces) and vice versa.  This is the lattice version of
Hodge duality.

Invariance of the theory under replacement of a diagram with its dual
diagram not only requires that both diagrams actually exist as Feynman
diagrams of that theory, but also that the weights of those two diagrams
be identical.  The symmetries of two dual diagrams are the same, so they
have identical combinatoric factors; thus we need to compare only the
products of their coupling constants.  Duality relates diagrams that not
only have different numbers of vertices, but also vertices with different
numbers of lines, so it relates the different n-point couplings.  Since any
theory has diagrams with loops with an arbitrary number of sides, the dual
diagrams will have vertices with an arbitrary number of lines, so the
potential must be nonpolynomial. 

It is simplest to consider vacuum graphs, and the duality of the
corresponding polyhedra (which approximate the sphere).  The
``watermelon" graph consisting of n propagators connecting two n-point
vertices is dual to a single loop with n two-point vertices.  From this we
find
$$ N^n (g_n)^2 = N^2 (g_2)^nâÜâ
	\left|{g_n\over N}\right| = \left|{g_1\over N}\right|^n $$
 which is sufficient to determine all couplings in terms of the one-point
coupling, up to signs.  (In particular, it tells us $g_1$ is nonvanishing.)  The
``daisy" graph, consisting of a single 2n-point vertex with n loops, is dual
to the ``pincushion" tree graph, consisting of one n-point vertex whose
legs are terminated by n one-point vertices.  This gives the relation
$$ N^{n+1}g_{2n} = N g_n(g_1)^nâÜâ
	{g_{2n}\over N} = {g_n\over N}\left({g_1\over N}\right)^n $$
 which determines all even-point vertices from lower-point ones: e.g.,
$g_2/N=(g_1/N)^2$, $g_4/N=(g_1/N)^4$.  Replacing one ``petal" of the
daisy with a ``pin" does the opposite on the dual graph, giving the relation
$$ N^{n+1}g_{2n+1}g_1 = N^2 g_{n+2}(g_1)^nâÜâ
	{g_{2n+1}\over N} = {g_{n+2}\over N}\left({g_1\over N}\right)^{n-1} $$
 which determines all odd-point vertices except $g_3$ from lower-point
ones.  Finally, the lollipop graph (loop with a stem) with a two-point
vertex on the loop is dual to the figure-eight graph with a two-point
vertex on one loop, so
$$ N^2 g_2 g_4 = N^3 g_1 g_2 g_3âÜâ
	{g_3\over N} = {g_4\over N}\left({g_1\over N}\right)^{-1} $$
 These diagrams are thus sufficient to solve for all vertices in terms of
the one-point vertex.  The unique solution for the n-point couplings is then
$$ g_n = N g^n $$
 in terms of a single coupling $g$.  

One might worry that duality for all Feynman graphs would
overdetermine these couplings, but the identity (from counting ends of all
lines)
$$ ÝnV_n = 2P - E $$
 for $V_n$ n-point vertices, $P$ propagators, and $E$ external lines shows
that $ÝnV_n$ is an invariant under duality (since $P$ and $E$ are).  Thus in
vacuum graphs $g$ always appears as $g^{2P}$:  While usually either the
total number of vertices $V$ ($=ÝV_n$) or the number of loops $L$ is used
to define the ``area" of the lattice, the duality invariant definition is to
use something proportional to their average, which follows here from the
usual $\hbar$-counting identity
$$ P - V = L - 1âÜâP = V+L-1 $$

The potential for the hermitian $NðN$ matrix field $Ä$ appearing in
$e^{-S}$ is then
$$ - N Ý_{n=1}^¥ \f1n g^n Ä^n = N¼ln (1 - gÄ) $$
 (The combinatoric factor is $1/n$ and not $1/n!$ because we consider
only planar diagrams, so lines from a vertex are ordered.)  This result can
also be obtained as a special case of the work of [6], where general
weights for both vertices and loops were considered, by restricting their
potential to be U($N$) invariant and have equal couplings for vertices and
loops.

This gives the action
$$ S = tr \int {d^D x\over (2¹)^{D/2}}\left[ üÄe^{-õ/2}Ä 
	+ N¼ln (1 - gÄ) \right] $$
 If we rescale $Ä£Ä/g$ we can associate the coupling with the kinetic
term:
$$ S' = tr \int {d^D x\over (2¹)^{D/2}}\left[ \f1{2g^2}Äe^{-õ/2}Ä 
	+ N¼ln (1 - Ä) \right] $$
 along with a rescaling of the functional measure.  This gives the above
counting of $g$'s in terms of propagators.  The form appropriate for the
$1/N$ expansion is obtained by further rescaling $g£g/åN$:
$$ S'' = N¼tr \int {d^D x\over (2¹)^{D/2}} \left[ \f1{2g^2}Äe^{-õ/2}Ä 
	+ ln (1 - Ä) \right] $$
 This model was considered in [7].  It is related to the Penner model [8],
except that here (the exponential of) the worldsheet cosmological
constant modifies the kinetic term rather than the potential.

Ü3.  DUALIZATION OF RANDOM MATRIX FIELDS

We now specialize to the simplest case, D=0.  Then
$$ S_Ä = tr \left[ üÄ^2 + N¼ln (1 - gÄ) \right] $$
 with the unitary ``color" symmetry
$$ Ä £ U_c Ä U_c^{-1} $$
 The logarithmic term suggests its derivation from bosonic ``quarks", in
addition to the scalar ``gluons" $Ä$. This gives a more natural-looking
form for the action that is real for all values of the fields:
$$ S_{ÆÄ} = tr \left[ üÄ^2 + Æÿ(1 - gÄ)Æ \right] $$
 for the complex ($NðN$) matrix $Æ$ with $N$ colors and $N$ flavors,
which gives the previous action as an exact result after Gaussian
integration. The quark field transforms under both color and flavor
symmetries
$$ Æ £ U_c Æ U_f^{-1} $$
 Unlike QCD, this $1/N$ expansion with gluons and quarks describes only
closed strings and not open, since in models with the number of flavors
proportional to the number of colors the topological expansion is identical
to that of pure Yang-Mills theory.  The action $S_{ÆÄ}$ actually somewhat
resembles 2D QCD [9] in that there is no gluon self-interaction.  

We can instead perform the $Ä$ integration on $S_{ÆÄ}$, which is now
also Gaussian, to find
$$ S_Æ = tr \left[ ÆÿÆ -üg^2 (ÆÿÆ)^2 \right] $$
 The quartic term has the wrong sign, just as in the usual $Ä^4$ matrix
model of strings.  Such models have been studied in [10,11], and shown to
give results similar to those of hermitian matrix models.

This action can in turn be derived from an action similar to $S_{ÆÄ}$, but
with $Ä$ replaced by a flavor field $$:
$$ S_{ƍ} = tr \left[ ü^2 +Æ (1 - g)Æÿ \right] $$
 where the new field transforms under only the flavor symmetry
$$  £ U_f  U_f^{-1} $$
 Finally, we reverse the initial step by integrating out $Æ$ to obtain an
action identical in form to the original one $S_Ä$:
$$ S_ =tr \left[  ü^2 + N¼ln (1 - g) \right] $$

The most interesting thing about these transformations is that they
explicitly perform a duality transformation on the matrix model action. 
To relate to the usual definition of T-duality in terms of Feynman graphs,
we examine the effects of these field transformations on the diagrams,
and observe that they explicitly perform a construction that replaces a
graph with the dual graph, where the original graph is drawn with color
lines, while the dualized graph is drawn with flavor lines.  As usual, each
propagator is drawn with a double line, each line representing an index on
the matrix, and the continuity of the lines representing the (unitary)
symmetries.  The planar graphs are those in which no lines cross, and all
external lines are on the outside of the graph.  (Actually, we here consider
vacuum graphs drawn on the sphere.)  The original graph is one following
from $S_Ä$.  The equivalent graph from $S_{ÆÄ}$ results from inserting a
(one-line) flavor loop into each vertex.  Eliminating $Ä$ to obtain $S_Æ$
has the effect of contracting the $Ä$ propagators to points.  Then
introducing $$ to obtain $S_{ƍ}$ causes $$ propagators to expand the
$Æ$ four-point vertices.  Thus the change from $S_{ÆÄ}$ to
$S_{ƍ}$ has replaced all $Ä$ propagators (double color lines) with $$
propagators (double flavor lines) pointing in the orthogonal direction. 
Finally, eliminating $Æ$ to obtain $S_$ simply shrinks all the color loops
to points.  Thus all color vertices have been replaced with flavor loops, all
color loops have been replaced with flavor vertices, and all color
propagators have been replaced with orthogonal flavor propagators.  So,
switching from color to flavor is exactly a duality transformation.

We can also consider more general models, using the same form of action
as $S_Æ$, with $N$ colors but with a different number $M$ of flavors
[10,11].  Such a theory is not duality invariant, but duality transforms a
modified $S_Ä$
$$ S_Ä = tr \left[ üÄ^2 + M¼ln (1 - gÄ) \right] $$
 for $NðN$ $Ä$ into a modified $S_$ that is identical in form to the
previous but with $$ $MðM$.  In particular, we can thus duality
transform this $S_Ä$ for the case $M=1$ to an $S_$ for a single variable
($1ð1$ matrix) $$.  Such ``vector" models have been studied in [12,11].
The double-scaling limit used for the Penner model [13] and its
generalizations kept $N-M$ fixed at a nonvanishing value, while
effectively taking the limit of our $g£¥$, while in our case we must fix
$N-M=0$ for duality invariance, and take a double-scaling limit where $g$
goes to a finite nonvanishing constant.

For D>0 an auxiliary term for $Æ$ as in $S_{ÆÄ}$ generates a $¶^D(0)$
coefficient for the logarithmic term, which violates duality, so the form of
the duality transformation will require some generalization.

ÜACKNOWLEDGMENTS

I thank Fabian E\ss ler for $N$ invaluable discussions. This work was
supported in part by the National Science Foundation Grant No.¼PHY
9309888.

\refs

£1 H.B. Nielsen and P. Olesen, \PL 32B (1970) 203;\\
	B. Sakita and M.A. Virasoro, \PR 24 (1970) 1146;\\
	G. 't Hooft, \NP 72 (1974) 461;\\
	F. David, \NP 257 [FS14] (1985) 543;\\
	V.A. Kazakov, I.K. Kostov and A.A. Migdal, \PL 157B (1985) 295;\\
	M.R. Douglas and S.H. Shenker, \NP 335 (1990) 635;\\
	D.J. Gross and A.A. Migdal, \PR 64 (1990) 127;\\
	E. Br«ezin and V.A. Kazakov, \PL 236B (1990) 144. 

£2 W. Siegel, \PL 134 (1984) 318;\\
        T.H. Buscher, \PL 194B (1987) 59, É201B (1988) 466. 

£3 D.V. Boulatov, V.A. Kazakov, I.K. Kostov, and A.A. Migdal,
        \NP 275 [FS17] (1986) 641;\\
	W. Siegel, \PL 252B (1990) 558. 

£4 F. David and R. Hong Tuan, \PL 158B (1985) 435.

£5 W. Siegel, Actions for QCD-like strings, Stony Brook preprint
	ITP-SB-96-1 (January 1996), hep-th/9601002.

£6 V.A. Kazakov, M. Staudacher, and T. Wynter, LPTENS preprints
	95-9, 95-24, 95-56, and 96-07 (hep-th/9502132, 9506174, 9601069,
	and 9601153).

£7 I.K. Kostov and M.L. Mehta, \PL 189B (1987) 118.

£8 R.C.Penner, ÓBull.Am.Math.Soc.Õ É15 (1986) 73,
 	ÓJ.Diff.Geom.Õ É27 (1988) 35;\\
	J. Harer and D. Zagier, ÓInvent. Math.Õ É85 (1986) 457.

£9 G. 't Hooft, \NP 75 (1974) 461.

£10 T.R. Morris, \NP 356 (1991) 703;\\
	J. Ambj\o rn, J. Jurkiewicz, and Yu.M. Makeenko, 
	\PL 251B (1990) 517.

£11 A. Anderson, R.F. Myers, and V. Periwal, \PL 254B (1991) 89.

£12 S. Nishigaki and T. Yoneya, \NP 348 (1991) 787;\\
	P. Di Vecchia, M. Kato, and N. Ohta, \NP 357 (1991) 495.

£13 J. Distler and C. Vafa, ÓMod. Phys. Lett.Õ ÉA6 (1991) 259.

\bye